\documentclass[acmsmall, screen]{acmart}

\makeatletter\def\@captype{table}\makeatother

\usepackage{multirow}
\usepackage{multicol}
\usepackage{enumitem}
\usepackage{wrapfig}

\AtBeginDocument{%
  }

\setcopyright{acmcopyright}
\copyrightyear{2018}
\acmYear{2018}
\acmDOI{XXXXXXX.XXXXXXX}

\acmConference[Conference acronym 'XX]{Make sure to enter the correct
  conference title from your rights confirmation email}{June 03--05,
  2018}{Woodstock, NY}
\acmPrice{15.00}
\acmISBN{978-1-4503-XXXX-X/18/06}

\renewcommand\footnotetextcopyrightpermission[1]{}

\begin{document}

\title{From Barrier to Bridge: The Case for AI Data Center/Power Grid Co-Design}



\author{Noman Bashir}
\affiliation{%
  \institution{Massachusetts Institute of Technology}
  \city{Cambridge}
  \state{MA}
  \country{USA}}
\email{nbashir@mit.edu}

\author{Rob Sherwood}
\affiliation{%
  \institution{SwitchDin}
  \city{Newcastle West}
  \state{New South Wales}
  \country{Australia}}
\email{rob.sherwood@gmail.com}

\author{Le Xie}
\affiliation{%
  \institution{Harvard University}
  \city{Cambridge}
  \state{MA}
  \country{USA}}
\email{xie@seas.harvard.edu}

\author{Minlan Yu}
\affiliation{%
  \institution{Harvard University}
  \city{Cambridge}
  \state{MA}
  \country{USA}}
\email{minlanyu@g.harvard.edu}

\renewcommand{\shortauthors}{Bashir, Sherwood, Xie, and Yu}

\begin{abstract}
For over a century, the electric grid has relied on a single statistical assumption: \emph{load diversity}, the principle that the uncorrelated demands of millions of small consumers produce a smooth, predictable aggregate.
AI training data centers break that assumption.
A single hyperscale training campus can draw power comparable to a mid-sized city, driven by one tightly synchronized job whose demand swings by hundreds of megawatts in seconds.
This paper argues that the resulting entanglement of compute and power infrastructure requires a shift from implicit coexistence to explicit co-development between the historically decoupled data center and electric power industries.
We introduce the distinct design principles, operational philosophies, and economic incentives of each sector, and show why their cultural and technical misalignment makes coordination difficult.
We identify key research directions, from joint capacity planning, multi-timescale control, a compute--power protocol stack, to market innovation, that must be pursued to power the future of AI sustainably and reliably.

\end{abstract}





\maketitle
\section{Introduction}
\label{sec:intro}

Electric and Internet cloud services are critical infrastructure that affect the lives of billions of people around the world. 
Although historically developed and operated in near-isolation, both sectors are now undergoing rapid transformations that increasingly intersect. 
The electric grid is being reshaped by large-scale electrification, ranging from the adoption of electric vehicles (EVs) to the transition from natural gas heating to electric heat pumps, while cloud computing is being reshaped by the explosive growth of artificial intelligence (AI) workloads~\cite{Shehabi:2024:DCEnergy}. 
These simultaneous transitions are unprecedented in scale and are beginning to stress each other in ways that neither system was originally built to handle.
In July 2024, a single transmission fault in Northern Virginia caused roughly 1.5~GW of voltage-sensitive data center load to trip offline within seconds, an event that NERC compared to the abrupt loss of a large nuclear plant and that prompted its first formal alert on large-load reliability risk~\cite{NERC:2025:StateOfReliability, GridStatus:2025:ByteBlackouts}.
For computer scientists, the implication is stark: decisions about job scheduling, checkpointing, and cluster design are no longer contained inside the data center; they now propagate onto the physical grid within milliseconds.

\begin{wraptable}{r}{3in}
    \vspace{-0.3cm}
    \scriptsize
    \centering
    \caption{Planned AI datacenter infrastructure spending for top companies with public reporting}
    \vspace{-0.3cm}
    \begin{tabular}{c|c|c}
    \hline
       \textbf{Company} & \textbf{Planned 2025 Investment} & \textbf{Sources} \\
       \hline
       Amazon & \$105 billion & \\
       Apple & \$100 billion & \\
       Microsoft  & \$80 billion & \cite{msft-datacenter-spend}\\
       Alphabet (Google)  & \$75 billion & \\
       Meta Platforms & \$65 billion & \\
       \hline
    \end{tabular}
    \vspace{-0.3cm}
    \label{table:ai_spend}
\end{wraptable}
On the cloud side, investment in AI infrastructure has reached historic levels (\autoref{table:ai_spend}). 
Publicly announced 2025 spending by five major technology companies is projected to exceed the annual GDP of 80\% of the world's countries, and total investment may double once private and speculative projects such as the Stargate are included.
The 2026 trajectory is steeper still: combined capex guidance from Amazon, Google, Meta, Microsoft, and Oracle is projected near \$600--700~billion, a roughly 50\% year-over-year jump that pushes hyperscaler capital intensity toward 45--57\% of revenue, a ratio historically seen only in utilities~\cite{IEA:2025:EnergyAndAI}.
Such rapid expansion reflects a profound shift in the demands placed on hyperscale data centers, driven by the ever-increasing computational requirements of modern AI training and inference workloads.

Meanwhile, the electric power industry faces its own structural shock. 
The transition from internal combustion to EVs alone is expected to add roughly 1,100 TWh of new annual load to global grids~\cite{IEA:2022:GlobalEVOutlook}. 
Additional electrification from heat pumps and industrial decarbonization compounds this challenge. 
Climate-driven reliability requirements, such as undergrounding transmission lines, further increase infrastructure costs. 
To accommodate these shifts while maintaining reliability, utilities must invest in new generation, storage, and transmission capacity as well as modernize control systems for inverter-based renewables. 
The International Energy Agency estimates that required global grid upgrades will exceed \$600 billion by 2030~\cite{IEA:2023:ElectricityGrids}.

These two trends, global electrification and AI growth, are increasingly colliding. 
Historically, cloud data centers were under 100 MW in size and constituted a small fraction of a grid operator's daily load~\cite{datacenter-ratings-semianalysis}. 
For perspective, the entire state of California uses 30–40 GW on a typical day~\cite{CAISO:2025:TodaysOutlook}; a traditional 100 MW datacenter represented only a few tenths of a percent of total demand. 
Their load patterns were slow, smooth, and tied to human activity, making them straightforward to integrate with modest planning effort and fit well within grid operators' load diversity assumptions.
Modern AI training campuses break this paradigm. They are projected to be 2--10~GW in electrical demand, 20--100$\times$ larger than traditional data centers, and represent 6--32\% of California's load per site~\cite{Meta:2024:RichlandParishDC, OpenAI:2024:StargateOracle}.
In aggregate, U.S. data center electricity use climbed from 58~TWh in 2014 to 176~TWh in 2023 and is projected to reach 325--580~TWh by 2028, at which point data centers could consume 6.7--12\% of total U.S. electricity~\cite{Shehabi:2024:DCEnergy}; the International Energy Agency projects global data center electricity consumption to more than double from 415~TWh in 2024 to roughly 945~TWh by 2030~\cite{IEA:2025:EnergyAndAI}.
Worse, AI workloads produce large, externally unpredictable changes in power consumption as jobs start, stop, checkpoint, or fail with sub-second timing~\cite{Choukse:2025:PowerStabilization}; grid reliability officials have publicly compared the resulting load profile to that of a steel mill, characterized by ``very fast, very large ramps''~\cite{UtilityDive:2025:ERCOT}.
This volatility directly violates the load diversity assumption that has anchored grid planning and operations for a century.
At the same time, the electric grid is highly regulated, regional in authority, slow to expand, and designed around physical inertia and long-term cost recovery. 
Internet cloud providers are unregulated, global, profit-maximizing enterprises that iterate on timescales of months. 
Their profit margins differ by more than 3$\times$: utilities operate under regulated returns on equity that averaged 9.7\% across 2025 U.S. rate cases~\cite{Fontanella:2025:RateCase}, while hyperscale cloud providers report operating margins of 39.5\% for AWS and 41.5\% for Microsoft's cloud segment in Q1 2025~\cite{AWS:2025:OperatingMargin, Azure:2025:OperatingMargin}.
These differences yield fundamentally incompatible operational timescales, risk models, planning horizons, and incentives (\autoref{table:clash}), setting the stage for deep cultural and structural frictions.

\begin{table}[t]
    \scriptsize
    \centering
    \caption{A comparative analysis of data center and electric grid paradigms.}
    \vspace{-0.35cm}
    \begin{tabular}{c|c|c}
    \hline
       \textbf{Attribute} & \textbf{Data Center (Hyperscale/AI)} & \textbf{Electric Grid (Utility)} \\
       \hline
       \textbf{Primary Incentive} & Speed-to-Market, ``success at all costs'' & Reliability, ``obligation to serve'' \\
       \textbf{Profit Model} & Unregulated, unbounded margins	& Regulated, capped margins \\
       \textbf{Planning Timescale}  & 5 years  & 10+ years \\
       \textbf{Operational Timescale}  & Sub-microsecond to real-time  & Minutes to days \\
       \textbf{Risk Appetite}  & High, can afford to make big mistakes  & Extremely low, cannot afford mistakes \\
       \textbf{Reliability Model} & Software-defined, ``RAID'' & Hardware-defined, ``Mainframe''\\
        & (resilience from unreliable parts)	 & (failure prevention) \\
       \textbf{Abstraction Model} & Software-defined, layered & Physics-defined, hierarchical \\
        & (OS / network / application) & (generation / transmission / distribution) \\
       \textbf{Geographical Scale} & Global & Strictly regional \\
       \hline
    \end{tabular}
    \vspace{-0.5cm}
    \label{table:clash}
\end{table}

Taken together, these developments drive the central claim of this paper:
\emph{Despite a century of engineering practice emphasizing loose coupling between critical infrastructures, the electric grid and AI training datacenters have become tightly intertwined, to the point that each has now become a bottleneck for the other.
To maintain reliability and enable continued scaling, the two systems must shift from implicit coexistence to explicit co-development.}

This paper articulates why such co-development is necessary, what makes it culturally and technically challenging, and where the opportunities lie. 
In particular, we identify the fundamental differences, in control timescales, economic incentives, regulatory oversight, and operational philosophies, that complicate coordination. 
At the same time, we highlight the underexplored opportunities created by datacenters' fast, programmable power electronics and the grid's large-scale buffering, suggesting new research directions in cross-layer control, market design, and planning that treat grid–datacenter interaction as a first-class systems problem.

The rest of the paper is organized as follows.
\autoref{sec:status-quo} reviews the load-diversity assumption that has governed grid operation for a century and explains why traditional data centers did not challenge it.
\autoref{sec:breakdown} characterizes how AI training workloads violate that assumption along four compounding dimensions: magnitude, second-to-second ramps, high-frequency oscillations, and extreme spatial concentration.
\autoref{sec:challenges} examines the cultural, technical, and market barriers that make co-development difficult.
\autoref{sec:research-directions} lays out four research directions, integrated planning, multi-timescale control, a compute--power protocol stack, and market innovation, that together define how cross-domain coordination can be built.
\autoref{sec:inference} looks beyond training to the emerging, qualitatively different challenge of large-scale AI inference.

\section{The Historical Status Quo: Electric Grid's Reliance on load diversity of Electricity Demand Across Customers}
\label{sec:status-quo}

For over a century, the operation of the electric grid has rested on a single statistical principle: \textbf{\emph{load diversity}}.
Grounded in the law of large numbers, load diversity holds that while individual loads, from residential refrigerators to industrial motors, are \emph{spiky} and \emph{unpredictable} in isolation, their fluctuations occur at different times.
Aggregated across millions of users, the peaks and valleys cancel, yielding an aggregate that is smooth, slow-moving, and manageable without knowledge of any individual customer's behavior.

\subsection{The Load Diversity Assumption}
Demand becomes even more predictable when combined with forecasts for external factors, such as weather or major sporting events. 
However, a critical distinction must be made between \textbf{\emph{correlation}} and \textbf{\emph{synchronization}}. 
Residential loads are often correlated by external factors; for example, a heatwave drives up air-conditioning use across an entire city, raising the grid's total demand.
Despite this correlation, these loads remain effectively unsynchronized at the second-by-second level: differing thermal inertia, building insulation, and thermostat hysteresis cause millions of AC compressors to cycle on and off in a staggered fashion.
This natural staggering is quantified by the \emph{diversity factor}~\cite{Willis:2004:PowerDistribution}, the ratio of the sum of individual maximum demands to the system's actual peak demand.
For residential distribution feeders, diversity factors typically range from~1.4 to~3.0~\cite{Willis:2004:PowerDistribution,NREL:2022:ResStock}, meaning that even during a coincident weather event the grid's peak is 30--70\% below the theoretical sum of individual peaks.
Without this statistical slack, utilities would face the prohibitive cost of sizing generation and transmission for a worst case in which every device activates simultaneously.

This operational paradigm simplifies grid management significantly, yielding a loosely coupled system in which utilities can focus on balancing aggregate supply and demand.

\subsection{Traditional Data Centers Do Not Break the Pattern}
Historically, even the largest data centers did not challenge this paradigm.
Pre-AI facilities were typically under 100~MW~\cite{Shehabi:2024:DCEnergy,datacenter-ratings-semianalysis} and exhibited predictable, slow-moving demand driven by the statistical aggregate of millions of independent end-user requests.
Their aggregate load varied on diurnal timescales of hours, not seconds, giving grid operators ample time to dispatch generation.
Consequently, planners integrated these facilities as large but well-behaved ``point loads.'' Siting them required careful planning for local transmission and generation capacity, but the \emph{nature} of their demand did not violate load diversity: they were simply large, predictable components of the statistical mix, not sources of system-destabilizing volatility.
\subsection{How Load Diversity Shaped the Grid's Architecture}
Load diversity produced a distinctive operational philosophy: the grid can, and should, remain \emph{agnostic} to what its customers do.
Because diversity guarantees predictable aggregate behavior, grid operators do not need to know the moment-to-moment energy needs of every load; they balance total generation against total demand.
This abstraction is structurally enforced by the electrical meter, a coarse-grained interface that records cumulative usage at low temporal resolution~\cite{bashir2021enabling}.
Utilities do not see which devices are drawing power or why; customers do not see how their electricity is generated or routed.
This bidirectional opacity is not a bug but a feature: it enabled the grid to scale by treating all users as independent contributors to a probabilistic aggregate.

The same assumption shaped sensing, control, and markets.
Telemetry evolved to sample at the substation or regional level on minute-to-hour timescales; central control loops operated on similar horizons.
Markets, rather than real-time coordination, became the primary tool for aligning supply and demand: since customer behavior was statistically predictable, day-ahead and real-time prices could schedule generation without tight synchronization across the utility--customer interface.
The grid functioned as a loosely coupled system in both engineering and institutional terms.

This century-old abstraction is now under pressure.
The assumption that all loads are small, uncoordinated, and slow-moving no longer holds for AI training clusters, which are large, synchronized, and rapidly variable.
Such loads do not simply raise the magnitude; they invalidate the principles on which load diversity itself depends.
Section~\ref{sec:breakdown} characterizes this break in detail.

\section{The Emerging Status Quo: AI Training Clusters Break Down Load Diversity}
\label{sec:breakdown}
The rise of hyperscale AI clusters marks a decisive break from the grid's historical equilibrium.
What was once a system stabilized by the statistical independence of millions of small users now faces a large city sized load, driven by a single training job run by a handful of engineers.
AI clusters do not just challenge grid operations; they invalidate the assumptions on which the grid was built.

\begin{table}[t]
    \scriptsize
    \centering
    \caption{Characteristics of three classes of data center. ``Power Predictability'' refers to the stability of electrical demand as seen by the grid: traditional and inference workloads vary with end-user activity, while AI training jobs are scheduled in advance but exhibit synchronized, high-amplitude power swings once running.}
    \vspace{-0.35cm}
    \begin{tabular}{c|c|c|c}
    \hline
        & \textbf{Traditional (CPU)} & \textbf{AI Training} & \textbf{AI Inference} \\
       \hline
       \textbf{Size} (\# of Servers) & 10s of thousands & 100s of thousands to millions & 10s of thousands  \\
       \textbf{Size} (\# of Sites) & Hundreds & A few dozen & Hundreds \\
       \textbf{Size} (Power Demand) & Up to 100~MW & 100s of MW to 10~GW & 100s of MW \\\hline
       \textbf{Power Predictability}  & Smooth, diurnal & Scheduled start but & Smooth, diurnal  \\
         & (user-driven) & synchronized, volatile & (user-driven) \\
       \hline
    \end{tabular}
    \vspace{-0.7cm}
    \label{table:dc_types_plan}
\end{table}

\subsection{AI Training is a Monolithic, Correlated Load}
AI training workloads violate the load-diversity assumption at its root.
\autoref{table:dc_types_plan} contrasts the three classes of data center along size and power-predictability axes:
a traditional data center aggregates millions of asynchronous user-driven requests, while an AI training cluster executes one massive computation in which tens of thousands of GPUs operate in lockstep.
The entire facility becomes a \emph{single correlated load}, not a statistical ensemble.
From the grid's perspective, the smooth probabilistic aggregate has been replaced by a deterministic, high-impact signal.

The closest historical analogs are heavy-industry facilities such as aluminum smelters and semiconductor fabs, which can consume several hundred MW and, in rare cases, approach 2.4~GW~\cite{CICS:2024:FabPower}.
Utilities have long managed these loads through bespoke interconnection agreements, wholesale-price exposure, and emergency shut-down clauses.
But even the largest industrial loads are slow-moving base load; they do not exhibit the synchronized, second-scale volatility of AI training.
AI training facilities combine the magnitude of a smelter with the switching behavior of a digital circuit, a combination the grid has never before encountered, and one that recent NERC reliability assessments identify as a distinct, emerging class of risk: NERC's 2025 Long-Term Reliability Assessment flags five major North American regions as ``high risk'' of resource inadequacy within five years (up from zero the prior year), and its September~2025 Large Load Industry Recommendation is the first formal NERC guidance addressing data-center interconnection risk specifically~\cite{NERC:2025:StateOfReliability,NERC:2025:LTRA,NERC:2025:LargeLoadIR}.

\subsection{Four Compounding Grid-Scale Challenges}
Treating an AI training cluster as an ordinary ``point load'' ignores the dynamics that make it unlike any prior industrial facility.
AI training introduces four compounding challenges: \emph{magnitude}, \emph{second-to-second ramps}, \emph{high-frequency oscillations}, and \emph{spatial concentration}.
Each would be destabilizing on its own; together, they constitute a qualitatively new reliability risk.

\begin{wraptable}{r}{3in}
    \scriptsize
    \vspace{-0.4cm}
    \centering
    \caption{Share of regional peak and average hourly demand represented by a single 10~GW AI training campus across five major U.S. balancing authorities. Values computed from~\cite{EIA:2025:GridMonitor}.}
    \vspace{-0.3cm}
    \begin{tabular}{lrrr}
    \hline
    Balancing  &  Peak Hourly &  Average Hourly & 10~GW Cluster\\
    Authority & Demand (MW) & Demand (MW) & (\% of peak)\\
    \hline
    CAISO  &  47{,}571 & 25{,}898 & 21.0\% \\
    ERCOT  &  85{,}544 & 54{,}381 & 11.7\% \\
    ISO-NE &  25{,}898 & 13{,}127 & 38.6\% \\
    NYISO  &  31{,}857 & 17{,}285 & 31.4\% \\
    PJM    & 160{,}560 & 94{,}507 &  6.2\% \\
    \hline
    \end{tabular}
    \label{tab:demand_stats}
\end{wraptable}
\noindent
\textbf{Magnitude.}
AI training campuses now reach 300~MW--10~GW, with 2~GW facilities already partially operational~\cite{OpenAI:2025:TenGWDC,CNBC:2025:AWSIndiana}.
\autoref{tab:demand_stats} quantifies the scale: a single 10~GW campus would represent 6--39\% of peak regional demand across the five largest U.S. balancing authorities.
The risk is not only steady-state capacity but sudden loss of load or massive synchronized surges, behaviors that even the largest industrial loads do not exhibit.

\vspace{0.05cm}
\noindent
\textbf{Second-to-second swings.}
The harder challenge is rate of change.
AI workloads exhibit load ramps of tens to hundreds of megawatts per second when thousands of accelerators synchronize at job start, enter or exit a checkpoint, recover from a fault, or hit a global barrier~\cite{Choukse:2025:PowerStabilization}.
Although the surge originates at the rack level, it propagates facility-wide because training jobs stay in lockstep: PDUs, substations, and the interconnection point all see the same synchronized step change.
This behavior bypasses the grid's natural averaging and outpaces primary frequency controls such as automatic generation control (AGC), which operate on minute-scale timescales.
The July 2024 Northern Virginia event, in which approximately 1.5~GW of voltage-sensitive data center load tripped offline within seconds, is the first regulator-documented instance of this failure mode at grid scale~\cite{NERC:2025:StateOfReliability,ESIG:2025:LargeLoads}.
Texas RE's director of reliability services has publicly compared AI data center load profiles to those of steel mills, characterized by ``very fast, very large ramps''~\cite{UtilityDive:2025:ERCOT}: a software event inside the data center has become a regional frequency disturbance.

\vspace{0.05cm}
\noindent
\textbf{High-frequency oscillations.}
Even during ``steady'' operation, large AI jobs induce second-to-millisecond oscillations caused by stragglers, pipeline bubbles, and collective-communication barriers.
When thousands of accelerators pause and resume together, power draw oscillates in unison.
These fast oscillations impose thermal and electrical stress on power electronics and transformers, cause voltage ripple and rapid current swings, and can excite resonant modes in rotating machinery; utilities have begun to flag this as an emerging reliability concern~\cite{ESIG:2025:LargeLoads}.
Traditional industrial loads do not produce internally synchronized millisecond-scale dynamics; AI workloads do so by design, accelerating fatigue on turbine shafts and other long-lead-time generation equipment.

\vspace{0.05cm}
\noindent
\textbf{Spatial concentration.}
Finally, the geographic concentration of AI data centers compounds every other effect.
If the same number of accelerators were spread across hundreds of sites, as in traditional cloud deployments, their independent workloads would partially cancel and load diversity would reappear.
Instead, hyperscale AI campuses pack hundreds of megawatts to multiple gigawatts of tightly synchronized load into a single substation.
Local feeders, transformers, and protection schemes, engineered for smooth industrial loads, are overwhelmed by the speed and amplitude of AI training dynamics.
The implication extends beyond the last mile: transmission lines have thermal limits, and absorbing a multi-gigawatt synchronized swing (\autoref{tab:demand_stats}) requires reinforced transmission to carry excess power out of the affected region as well as higher-voltage connections into it, so large portions of the broader network must be upgraded, not only the path from generation to the data center.

\subsection{A Paradigmatic Shift}
These four dimensions collectively invalidate the grid's century-old abstraction.
Maintaining reliability now requires explicit coordination across timescales, from millisecond-level inverter response to multi-year capacity planning, rather than the loose coupling that load diversity used to permit.
The shift is paradigmatic: AI training customers can no longer be treated as stochastic noise within a probabilistic aggregate; they must be treated as dynamic actors whose software-level decisions have physical, grid-scale consequences.
As the next section argues, this shift confronts deep cultural, technical, and market misalignments between the two sectors.

\section{The Case for Co-Development and its Cultural and Technical Challenges}
\label{sec:challenges}
Unlike prior industrial loads, AI data centers are equipped with fast, programmable power electronics and workload-level controls that could, in principle, help stabilize the grid.
Facilities already deploy multi-megawatt UPS inverters, battery banks, DC buses, and job-level schedulers capable of shaping electrical demand from milliseconds to hours.
But despite the local capability, the two sectors remain culturally, technically, and institutionally unable to coordinate.
Before dissecting those gaps, we first address a natural objection: if data centers already have the right knobs, and grids already have the right capacity, why can't each side solve its problems independently?

\subsection{Why Local Solutions Fall Short}
The data center side already has a rich set of flexibility mechanisms, because most of them were built for reliability, not for the grid.
Uninterruptible power supplies sized for utility ride-through, dual-fed PDUs, software-defined failure domains, and checkpoint--restart machinery were all engineered to let the facility survive disturbances without data loss or downtime.
Recent work shows these same mechanisms can shape electrical demand: Microsoft's power-stabilization system uses UPS inverters and workload throttling to smooth sub-second load swings~\cite{Choukse:2025:PowerStabilization}, and follow-on industry and academic work has extended the idea to hour-scale shifting and grid-interactive operation without new hardware~\cite{Radovanovic:2022:CarbonAware,nvidia-dc-demo,Lin:2025:GridInteractive}.
The knobs exist.

But local actuation alone does not close the gap.
UPS batteries sized for seconds of ride-through cannot absorb multi-hour load imbalances, and no amount of in-facility smoothing changes the fact that a 10~GW campus still draws 10~GW that must come from somewhere (\autoref{tab:demand_stats}).
The recent interest in behind-the-meter generation illustrates the point.
An on-site small modular nuclear reactor would address magnitude and spatial concentration by bringing generation to the point of load, but nuclear is an inflexible baseload resource with ramp rates measured in hours, precisely the wrong technology for second-to-second swings and millisecond oscillations.
A grid-independent data center paired with a dedicated reactor must absorb its own fast volatility internally, which is exactly what the interconnected grid was invented to provide; self-isolation forfeits bulk-system inertia and statistical buffering, leaving the facility more exposed, not less.
Neither side has all the ingredients: data centers have fast actuation but limited bulk capacity; grids have bulk capacity but slow actuation.
The remainder of this section examines three kinds of misalignment, cultural, technical, and market, that keep the two sides from combining what they have.

\subsection{Cultural and Institutional Misalignment}
The two industries are guided by fundamentally different philosophies of reliability, risk, and time.
Utilities are regulated, mission-driven entities whose social contract demands stability, fairness, and universal service; hyperscale operators are competitive, globally distributed enterprises rewarded for speed, scale, and innovation.
The resulting clash is captured by two opposing reliability models: the grid's \emph{mainframe} approach, which hardens every component to prevent failure, versus the data center's \emph{RAID} approach~\cite{Patterson:1988:RAID, Luiz:2019:DCBook}, which assumes frequent failures and routes around them in software.
The RAID model does not transfer cleanly to power systems: nothing in a high-voltage grid is inexpensive enough to treat as disposable, and transmission and generation components have multi-year replacement lead times rather than next-day RMA cycles.

Planning horizons amplify the divide.
Utilities invest on decade-long horizons under regulatory review and public cost recovery; hyperscale operators iterate on five-year build cycles driven by shifting workloads, accelerator generations, and market competition.
By the time a transmission project is approved, the workload it was meant to serve may have migrated or changed entirely.
Economic incentives compound this: utilities operate under regulated returns on equity that averaged 9.7\% across 2025 U.S. rate cases~\cite{Fontanella:2025:RateCase}, while the leading hyperscale cloud segments reported operating margins of 39.5\% (AWS) and 41.5\% (Microsoft) in Q1 2025~\cite{AWS:2025:OperatingMargin, Azure:2025:OperatingMargin}.
One side prioritizes equity and long-lived assets; the other optimizes for model timelines and cost per delivered token.

This misalignment paralyzes co-development because it destroys the trust that joint engineering requires.
A utility cannot rate-base a billion-dollar transmission upgrade against a cloud roadmap that may be rewritten in two quarters; a hyperscaler cannot stake a training schedule on a ten-year interconnection study.
Co-development requires aligned incentives, synchronized planning, and shared risk, none of which the existing institutional cultures naturally produce.

\subsection{Technical and Architectural Incompatibilities}
The two systems' control architectures were never designed to interoperate.
The grid evolved as a hierarchical, physics-constrained control system with coarse telemetry and slow feedback loops: seconds for primary frequency response, minutes for automatic generation control, and hours for unit commitment.
Data centers run deeply layered digital loops spanning microseconds (voltage regulation, power control), milliseconds (job scheduling), and seconds (autoscaling, placement).
The two sides therefore observe the environment at vastly different temporal resolutions, and a disturbance that is instantaneous inside the data center is visible to the grid after the fact.

This mismatch is what turns a rack-level synchronization event into a grid-scale disturbance.
Because the grid aggregates at the interconnection point rather than at the rack, thousands of synchronized accelerators appear as a single unified step, not as independent micro-events that could cancel out.
Meanwhile, the interconnection process itself operates on the grid's clock: each new facility triggers a bespoke chain of feasibility studies, power-flow analyses, protection reviews, and transmission-upgrade planning that can take 3--7 years~\cite{Farney:2025:PowerPlay}.
That cadence was tolerable when data centers were 50--100~MW; with multi-gigawatt AI campuses, the per-site study process cannot keep up with 2--5 year build cycles.
As noted in~\autoref{sec:challenges}, the fast-actuation hardware needed to coordinate already exists inside the data center, but its telemetry is not exposed, its actuation is proprietary, and its reliability envelopes are scoped inward only.
There is no common visibility layer, no standard interface for inverter coordination, and no shared disturbance model, so both sides remain blind at the moment coordination matters most.
\subsection{Market and Governance Barriers}
Current economic and regulatory structures actively discourage collaboration.
Tariff frameworks often socialize the cost of interconnection and reliability upgrades across all ratepayers~\cite{Farney:2025:PowerPlay}, effectively subsidizing hyperscale customers rather than internalizing the volatility they introduce.
More fundamentally, the grid's primary demand-side levers are calibrated for a different kind of load.
Typical demand-response programs pay \$30--\$50 per kW-year~\cite{ISONE:2025:AMR}, and spot prices are capped near \$5,000/MWh~\cite{Potomac:2025:ERCOT}.
For a hyperscaler, stalling a frontier-model training run during a reliability event means delaying a product whose future revenue is measured in billions and whose time-to-market window is measured in months; a few thousand dollars of capacity payments is a rounding error.
The mismatch effectively disables the grid's main economic tool for modulating load.

Governance widens the gap further.
Utilities operate under public oversight, formal rule-making, and cost-of-service regulation; hyperscale operators iterate privately, globally, and at will.
No shared body, protocol, or institutional venue exists today for approving joint investments, allocating reliability risk, or enforcing mutual obligations, so even when cooperation is in both parties' interest, there is no machinery to commit to it.
Utilities cannot rate-base transmission against a counterparty that may change its mind in two quarters; data center operators cannot invest in grid-supportive capabilities without predictable compensation and interconnection rules.

A recent analysis from Duke's Nicholas Institute argues that, if large loads can curtail for a small number of hours per year, the existing U.S. grid could absorb roughly 100~GW of new demand without new generation~\cite{Norris:2025:Rethinking}.
The analysis is optimistic and depends on real curtailment capability; our point is precisely that today's AI training clusters, because of the opportunity-cost gap above, have no incentive to expose the flexibility that the headroom assumes.
Co-development is the mechanism by which that latent flexibility becomes real: it aligns incentives, builds the interfaces, and closes the institutional gap so that the grid's physical capacity and the data center's programmable actuation can reach each other.

\section{Future Research Directions}
\label{sec:research-directions}
Closing the gaps identified in~\autoref{sec:challenges} requires more than ad hoc fixes; it demands a redesign of how data centers and grids plan, operate, and exchange information.
We organize the research agenda into four directions: integrated planning, multi-timescale control, a layered \emph{compute--power protocol stack} that unifies them, and market innovation.
Fifteen years ago, Keshav and Rosenberg argued that Internet concepts, layered abstractions, feedback, and standard interfaces, could help smarten the grid~\cite{Keshav:2010:InternetConcepts}; AI training makes that call non-optional.

\vspace{0.05cm}
\noindent
\textbf{1 -- Integrated Planning.}
Today's interconnection process is reactive and transactional: the grid responds to each new siting request with an individual feasibility study, typically 3--7 years long (\autoref{sec:challenges}).
The goal is to replace this with proactive, joint planning in which projected computing demand feeds directly into resource-adequacy studies and transmission build-out.
Concretely, ``aligning'' the data center's 5-year build cycle with the grid's 10-year planning cycle means that the two sides agree on synchronized decision gates: every five years both sectors commit to a shared forecast of aggregate regional compute demand, accelerator-mix assumptions, and expected curtailment behavior, and the grid's corresponding resource-adequacy plan is sized accordingly.
Research challenges include: (i)~\emph{long-horizon capacity forecasting}, distinct from the minute-to-hour operational load forecasting addressed under Direction~2, that projects aggregate AI compute demand 5--10 years out under scenarios for model scaling, training-to-inference transition, and accelerator efficiency trends; (ii)~\emph{cross-domain scenario analysis} that couples AI roadmaps to grid stress events such as extreme weather and correlated load-loss incidents of the kind documented by NERC~\cite{NERC:2025:StateOfReliability}; and (iii)~\emph{institutional decision mechanisms}, such as joint reliability risk assessments and ``compute IRP'' processes modeled on the utility integrated resource planning framework, that give both sides a formal venue for binding, forward-looking commitments.

\vspace{0.05cm}
\noindent
\textbf{2 -- Multi-Timescale Control.}
Operational coordination must span orders of magnitude in time, from millisecond voltage stabilization to minute-level economic dispatch.
As~\autoref{sec:challenges} noted, the individual building blocks already exist (Google's carbon-intelligent computing~\cite{Radovanovic:2022:CarbonAware} at hour scale, Microsoft's and NVIDIA's power stabilization~\cite{Choukse:2025:PowerStabilization,nvidia-dc-demo} at sub-second scale, EPRI's DCFlex and Flex MOSAIC framework~\cite{EPRI:2025:FlexMOSAIC} on the utility side, and recent work showing that workload orchestration alone suffices for many grid-interactive behaviors~\cite{Lin:2025:GridInteractive}); what is missing is the coordination plane that composes them.
Open research problems include: (i)~programmable \emph{hybrid UPS/DC-bus buffers} that compose battery response at seconds with supercapacitor response at milliseconds; (ii)~\emph{secure, standardized telemetry} that exposes rack- or substation-level state to both sides (IEEE 2030.5 is a candidate starting point); and (iii)~\emph{cross-domain simulation platforms} that couple accelerator-level workload traces to grid transient-stability models, enabling pre-deployment analysis of new AI jobs before they hit the interconnection point.

\vspace{0.05cm}
\noindent
\textbf{3 -- A Compute--Power Protocol Stack.}
Directions~1, 2, and 4 naturally fall into three layers, and we argue that treating them as such, rather than as independent efforts, is the paper's core research proposition.
At the \emph{data layer}, shared telemetry and forecasts must cross the utility--customer boundary in both directions (breaking the century-old opacity of the electrical meter).
At the \emph{control layer}, inverter dynamics, workload schedulers, and grid regulation controls must co-optimize across millisecond-to-hour horizons.
At the \emph{economic layer}, tariffs, flexibility markets, and cost-causation rules must sustain the cooperation over decades.
The analog is Internet peering: independent ASes coordinate through standard interfaces (BGP, settlement-free peering agreements, SLAs) without unifying their internal operations or ownership; the grid--data center relationship can evolve similarly, with reliability redefined as a shared, emergent property of the joint stack rather than the sole responsibility of either side.
Research priorities include defining the minimum interface contracts at each layer, proposing regulatory sandboxes where new layer combinations can be tested without full regional commitment, and developing conformance testing suites analogous to those that enabled the Internet to scale.

\vspace{0.05cm}
\noindent
\textbf{4 -- Market and Economic Innovation.}
Current market structures misprice volatility and obscure responsibility.
Two complementary innovations are needed.
First, \emph{cost-causation tariffs} would charge customers in proportion to the incremental infrastructure and reserve capacity their usage requires, rather than socializing those costs across all ratepayers~\cite{Farney:2025:PowerPlay}; this internalizes externalities and creates transparent incentives for grid-friendly siting and workload design.
Second, \emph{flexibility markets} must value the latent controllability that AI training already possesses: jobs can be paused, checkpointed, and restarted, and geographically replicated training runs can shift across regions at hour timescales.
Candidate market products include ramp-rate limits priced as an ancillary service, deferred-compute contracts that pay hyperscalers to delay non-urgent training windows, and third-party ancillary-service purchases in which data center operators buy additional grid reserve capacity in exchange for relaxed ramp constraints.
Recent analyses suggest the prize is large: up to roughly 100~GW of new flexible load may fit within existing U.S. grid headroom if curtailment of a small number of hours per year is exposed to the market~\cite{Norris:2025:Rethinking}.
Making that prize real requires pricing mechanisms that close the two-orders-of-magnitude gap between traditional demand-response payments~\cite{ISONE:2025:AMR,Potomac:2025:ERCOT} and the opportunity cost of frontier-model training, so that exposing flexibility becomes a rational business decision rather than a charitable act.

\section{Inference: The Next Frontier of Grid Interaction}
\label{sec:inference}

This paper has focused on AI training because that is where the load-diversity violation is sharpest today.
But a second shift looms: AI inference, serving deployed models to users, is growing rapidly and has qualitatively different electrical behavior from training.
Training consumes capital while inference generates revenue~\cite{Patterson:2022:CarbonFootprint}, so every hyperscale business plan assumes that inference volume and its energy footprint will eventually rival or exceed training's; industry disclosures already indicate a substantial and growing share of accelerator revenue and deployed-compute time~\cite{Luccioni:2024:PowerHungry}.
Co-development therefore requires anticipating how inference loads, smaller per site but more numerous, geographically distributed, and coupled to end-user activity, will interact with the grid.

\vspace{0.05cm}
\noindent
\textbf{Inference Today: Small Per Site but Structurally Different.}
Inference currently accounts for a minority of AI compute energy, and its physical footprint looks unlike a training campus.
Inference fleets comprise many smaller deployments, typically 10--30~MW each, embedded in colocation facilities and metro-scale clusters connected at distribution voltages rather than high-voltage transmission lines~\cite{Luccioni:2024:PowerHungry}.
From the grid's perspective, these sites resemble commercial or light-industrial customers rather than heavy industry.
Because inference serves real-time user requests, its demand exhibits diurnal, weekly, and event-driven variability that reflects human activity and product release cycles, not synchronized software barriers.
Operators manage this volatility with predictive autoscaling and cross-region traffic steering, but these mechanisms produce short-timescale aggregate power ramps of their own: when millions of users open the same AI assistant at the start of a workday, inference fleets scale up together.
So far the aggregate grid impact has been modest; at projected scale it will not be.

\vspace{0.05cm}
\noindent
\textbf{Why Inference Will Eclipse Training.}
Inference growth is not optional; it is required by the business model.
Hyperscalers operate with gross margins in the 60--90\% range, and sustaining those margins as training capex compounds requires inference revenue to grow proportionally faster, translating to more compute-hours and more energy per deployed model~\cite{Patterson:2022:CarbonFootprint,Luccioni:2024:PowerHungry}.
Unlike training, however, inference is \emph{heterogeneous} in latency and reliability requirements: real-time services (conversational assistants, autonomous driving) need millisecond response, while offline batch services (video transcription, document summarization, embedding computation) can be deferred or degraded with little user impact.
This heterogeneity is the key new variable; the research question is how to quantify and expose the flexibility without violating service-level objectives.

\vspace{0.05cm}
\noindent
\textbf{The Shape of Future Inference Fleets.}
The spatial configuration of inference is not yet settled and will shape its grid interaction.
Three trajectories are plausible.
In an \emph{edge-proximity} scenario, user-facing latency dominates and inference migrates toward the network edge, into CDNs, metro data centers, and on-premise microclouds~\cite{Satyanarayanan:2017:EdgeComputing}; grid interactions shift from a few large interconnections to thousands of small ones distributed across urban load pockets.
In a \emph{data-gravity} scenario, large shared datasets and proprietary embeddings hold inference co-located with training campuses, preserving concentration and reinforcing existing transmission bottlenecks.
In the most likely \emph{hybrid} scenario, latency-critical inference runs near users while compute- or data-heavy inference runs in regional hubs, producing fine-grained diurnal cycles at the edge overlaid on longer cycles at the core.
These patterns have very different relationships to load-diversity: a geographically distributed inference footprint partially \emph{restores} load diversity, because many independent sites serving heterogeneous user populations and time zones let fluctuations cancel statistically again, while centralized inference preserves the concentration and volatility problems of training.

\vspace{0.05cm}
\noindent
\textbf{Research Opportunities.}
Inference inherits parts of the research agenda from~\autoref{sec:research-directions} and adds problems of its own.
Most of the protocol-stack, multi-timescale control, and market-innovation work applies with little modification, but the aggregation and spatial-diversity questions are unique to inference.
We highlight three directions not covered elsewhere:
\begin{itemize}[leftmargin=*, topsep=0cm, itemsep=0pt]
    \item \textbf{Quantifying inference elasticity.} Develop empirically grounded models that map the latency--cost--reliability surface of real inference services and estimate how much flexibility can be exposed to grid coordination without violating SLOs.
    \item \textbf{Aggregation and restored diversity.} Study how fleets of 10--30~MW inference sites aggregate statistically in space and time, and under what geographic distributions the load-diversity property of the pre-AI grid can be partially recovered.
    \item \textbf{Grid-aware scheduling at planetary scale.} Extend carbon-aware workload placement~\cite{Radovanovic:2022:CarbonAware} to full grid-stability signals, so that routing, autoscaling, and replica placement react in real time to frequency, voltage, and local reserve conditions, not just to carbon intensity.
\end{itemize}

Inference both continues and diverges from the training story: it shifts the problem from stabilizing a single monolithic load to coordinating millions of semi-elastic ones, which is simultaneously harder (many more control points) and easier (restored statistical diversity).
Its current footprint is modest; at scale, it will be the next frontier for grid--data center co-development.

\bibliographystyle{ACM-Reference-Format}
\bibliography{paper}











\end{document}